\documentclass{jltp}

\usepackage{graphicx} % uncomment this line to include the graphicx package

\title{Landau Damping of Spin Waves in Trapped Boltzmann Gases}

\author{R.J. Ragan, W.J. Mullin$^*$, and E.B. Wiita}

\address{Physics Department \\University of Wisconsin at Lacrosse, La Crosse, WI 54601, USA\\
$^*$Physics Department \\Hasbrouck Laboratory, University of Massachusetts, Amherst, MA 01003}

\runninghead{R. J. Ragan and W.J. Mullin}{Landau Damping of Spin waves in Trapped Boltzmann Gases}

\begin{document}
\maketitle
\begin{abstract}
A semiclassical method is used to study Landau damping of transverse pseudo-spin waves in harmonically trapped ultracold gases in the collisionless Boltzmann limit.
In this approach, the time evolution of a spin is calculated numerically as it travels in a classical orbit through a spatially dependent mean field. 
This method reproduces the Landau damping results for spin-waves in unbounded systems obtained with a dielectric formalism. 
In trapped systems, the simulations indicate that Landau damping occurs for a given spin-wave mode because of resonant phase space trajectories in which spins are \lq\lq kicked out" of the mode (in spin space).
A perturbative analysis of the resonant and nearly resonant trajectories gives the Landau damping rate, which is calculated for the dipole and quadrupole modes as a function of the interaction strength. The results are compared to a numerical solution of the kinetic equation by Nikuni {\it et al}.\end{abstract}

\section{INTRODUCTION}

Recent experiments at JILA\cite{ref1}$^,$\cite{ref2} have demonstrated transverse spin waves in the pseudo spin dynamics of the hyperfine levels of trapped ultra-cold $^{87}$Rb atoms in the nondegenerate regimes. As in spin-polarized Boltzmann gases, an exchange mean-field supporting collective spin behavior can occur when the thermal de Broglie wavelength exceeds the effective range of the interaction between colliding atoms\cite{ref3}$^,$\cite{ref4}.  Numerical simulations of the appropriate kinetic equations with JILA parameters have been shown to reproduce the observed pseudo-spin oscillations.\cite{ref5}  Using a moments method, Nikuni {\it et al}.\cite{ref6} obtained analytic expressions for the spin-wave frequencies and damping rates of the dipole and quadrupole modes in a trapped Boltzmann gas. The results were found to be in excellent agreement with a one dimensional numerical solution of the kinetic equation, except at intermediate densities, where Landau damping was thought to be important. However, an estimate using the Landau damping result for a homogeneous system\cite{ref7} yielded a rate that was order of magnitude too large.

One of the disadvantages of the moments method is that the truncated form of the distribution functions do not account for the mean field coupling of the collective modes to higher excitations that give rise to Landau damping. Indeed, Landau damping usually requires a pole in the spin distribution function.\cite{ref8} A trapped system poses another problem in that the excitation spectrum is discrete, whereas Landau damping calculations usually involve integration over a continuous spectrum. One method that is often used is to look at elongated clouds and treat the long axis of the trap as infinite,\cite{ref9} but the applicability of this approximation to low lying states is questionable. In the following we use a semi-classical approach to extend the moments method to account for Landau damping in trapped systems. 

\section{HOMOGENEOUS SYSTEMS}
Before considering trapped systems, we review Landau damping in a homogeneous system. We start with the kinetic equation for the normalized transverse spin distribution $s^+(z,p)$ in the collisionless regime, linearized for small tip angles
\begin{equation}
\partial_t s^+(z,p)=-(p/m)\partial_zs^+(p,z)
-i\Omega[s^+(z,p)-S^+(z)]
\label{ke}
\end{equation}
where $\Omega$ is the mean-field frequency, $S^+(z)=\int s^+(z,p) n(p)\,dp$ and $n(p)=(2\pi mk_BT)^{-1/2}\exp[-p^2/(2mk_BT)]$ is the normalized distribution in momentum space. Now substituting traveling wave solutions of the form $s^+(z,p)=s^+(p)\exp[i(qz-\omega t)]$, and solving for $s^+(p)$ we get
\begin{equation}
s^+(p)=\frac{\Omega S^+}{qp/m+\Omega-\omega} 
\label{mp}
\end{equation}
Integrating over $p$ and taking into account the pole at $p_L=(m/q)(\omega-\Omega)$
we get the same result obtained in Ref.\onlinecite{ref7} using an RPA approximation:
\begin{equation}
\omega=-\frac{k_BT}{\Omega m}q^2-i\frac{\Omega^2}{q}\sqrt{\frac{\pi m}{2 k_B T}}\exp\left[-\frac{1}{2}\frac{\Omega^2}{q^2}\frac{m}{k_B T}\right]
\label{omega}
\end{equation}

\section{SEMICLASSICAL APPROACH}
The above calculation of the damping rate is direct and simple, but it cannot be easily adapted to trapped systems. We now recalculate the above damping rate for the unbounded case using a semiclassical approach. In the next section we adapt this approach to a numerical method for trapped systems. 

If we consider the small transverse component of a spin as it travels along the helix, we find that its equation of motion is given by $d\vec{\sigma}(t)/dt= \Omega \vec{\sigma}(t) \times \vec{S}[z(t)]$. In linearized form this becomes
\begin{equation}
\dot{\sigma}^+=-i\Omega\left( \sigma^+-S^+[z(t)]\right)=-i\Omega\left( \sigma^+-e^{iqz(t)}e^{-i\omega_q t}\right)
\label{mdot}
\end{equation}
where in the collisionless regime, $\omega_q=-{k_BT}/{\Omega m}q^2$ and $z(t)=pt/m$. The solution is given by
\begin{equation}
\sigma^+(t)=2i\Omega \frac{\sin [(qp/m + \Omega- \omega_q)t/2]}{(qp/m + \Omega- \omega)}e^{i(qp/m - \Omega- \omega_q)t/2} 
\label{m}
\end{equation}\
Obviously, we have resonance at $p_L \equiv (\omega_q-\Omega)m/q$ in which case we have $|\sigma^+(t)|=\Omega t$. Otherwise, $\sigma^+$ is oscillatory, with an amplitude $2m\Omega/q(p-p_L)$ and a beat frequency $\omega_{beat}=q(p-p_L)/m$.  We note that in the absence of damping the amplitude of $\sigma^+\sigma^-$ is constant. Thus, the derivative of $\sigma^+\sigma^-$ gives the Landau damping rate. Using Eq.\ref{m} we have 
\begin{equation}
\lim_{t\rightarrow\infty} |\sigma(t)|^2/t=2 \pi \frac{m\Omega^2 }{q} \delta (p - (\omega_q- \Omega)m/q)
\label{GR}
\end{equation}
The contribution to the Landau damping of $s^+$ is half this value.\cite{ref9} To find the total landau damping rate we integrate over the $p$ distribution according to  
\begin{equation}
\gamma_L=\pi \frac{m\Omega^2 }{q} \int   
n(p) \delta (p - p_L)\, dp
\label{Ldamptot}
\end{equation}
which gives, assuming $\omega \ll \Omega$, the same result as Eq.\ref{omega}.

We now outline a numerical procedure to find the Landau damping rate that can be generalized to trapped systems. First we search for the resonant momentum $p_L$. We then measure the slope of the growth of $|\sigma^+(t)|$ at resonance to get $\Omega_{eff}$. Finally we find $q_{eff}$ from the beat frequency exhibited by $s^+(t)$ when $p$ is slightly off resonance. According to Eq.\ref{m} we have $q_{eff}(p-p_L)T_{beat}/m=2\pi$ or 
\begin{equation}
q_{eff}=\frac{2\pi m}{(p-p_L)T_{beat}}
\label{beat}
\end{equation}

\section{TRAPPED SYSTEMS}
We restrict our attention to elongated (quasi-1d) clouds. For a trapped system the trajectories through phase space are no longer straight lines of constant $p$, but ellipses of constant energy $E=p^2/2m +m\omega_z^2z^2/2$, where $\omega_z$ is the axial trap frequency. Eq.\ref{mdot} generalizes according to 
\begin{equation}
\dot{\sigma}^+=-i\Omega(z)\left( \sigma^+-S^+(z,t)\right)
\label{mdot3}
\end{equation}
with $z(t)=\sqrt{2E/(m\omega_z^2)} \cos(\omega_z t)$ and $\Omega(z)=\Omega_0 n(z)/n_0$, where $\Omega_0$ is radially averaged exchange frequency\cite{ref4} and $n(z)/n_0=\exp(-m\omega_z^2z^2/2k_BT)$ is the ratio of the density of the cloud to its peak value.  For the transverse magnetization we use as an approximation the results from the moments method of Nikuni {\it et al.}\cite{ref6} for the dipole and quadrupole modes:
\begin{eqnarray}
S_1^+(z)&=&c_1ze^{-i\omega_1t}\\ 
S_2^+(z)&=&c_2(z^2-1)e^{-i\omega_2t}
\label{moments}
\end{eqnarray}
where the frequencies are given by $\omega_j=-\Omega_0/\sqrt{8}+\sqrt{\Omega^2/8+j^2\omega^2_z}$ and the normalization constants $c_j$ are determined from the condition
\begin{equation}
1=\int|s^+(z,p)|^2n(z,p)dzdp
\label{norm}
\end{equation}
where $n(z,p)=\omega_z/(2\pi k_BT)\exp(-E(z,p)/kT)$. The Landau damping rate is then given by 
\begin{eqnarray}
\gamma_L&=&\frac{1}{2}\int\frac{|\sigma^+(t)|^2}{t} n(z,p)dzdp \nonumber\\ 
&=&\pi \frac{m\Omega_{eff}^2 }{q_{eff}} \int \sqrt{\frac{2E}{m}}\frac{1}{k_BT}e^{-E/k_BT}\delta
(E-E_L)dE \nonumber\\ 
&=&\pi\frac{m\Omega_{eff}^2 }{q_{eff}}\sqrt{\frac{2E_L}{m}}\frac{1}{k_BT}e^{-E_L/k_BT}
\label{ldamp2}
\end{eqnarray}

To calculate the Landau damping rate we use the numerical procedure described in the previous section. Solutions of Eq.\ref{mdot3} (See Fig.1) show the same sort of resonant behavior as the exact solution for the homogeneous system. However, the effective parameters for the trapped system are very different from those of the homogeneous system. For $\Omega_0=2\omega_z$ in the quadrupole mode we find a much smaller effective value $\Omega_{eff}=0.38 \omega_z$. For the same $\Omega_0$ we get $q_{eff}=0.237z_T^{-1}$ where $z_T=(k_BT/m\omega_z^2)^{1/2}$ is the thermal cloud size. This is much smaller than an estimate in which the distance between the nodes is taken to be a half a wavelength, giving $q=1.57\omega_z$. (Not surprisingly, the dipole mode has a smaller value $q_{eff}=0.158z_T^{-1}$.) Again for $\Omega_0=2\omega_z$, in the trapped system we find $E_L=4.36k_BT$ whereas the argument of the exponential in Eq.\ref{omega} gives 0.811.  Fig.2a shows the Landau damping rate calculated from Eq.\ref{ldamp2}. Fig.2b shows the collisional damping rate calculated via the moments method\cite{ref6} as well as the sum of the Landau and collisional damping rates. For comparison the numerical solutions of the kinetic equation of Nikuni {\it et al.}\cite{ref6} are also shown. Although the location of the Landau damping peaks are correct and they are of the right order of magnitude, there are qualitative differences with the numerical solutions. The peak Landau damping rate for the dipole mode is too large, whereas it is too large for the quadrupole at high densities.
\begin{figure}
\centering
\includegraphics[width=3.0in]{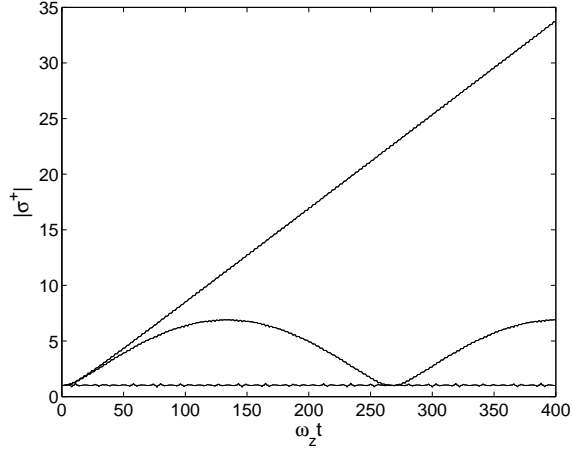} 
\caption{Time evolution of the transverse component of a spin as it moves through the dipole mode for $\Omega_0=3.5\omega_z$ at: the resonant energy $E=E_L$ (straight line), slightly off resonance showing beats, and far from resonance.}
\label{fig:one}
\end{figure}
\begin{figure}
\centering
\includegraphics[width=4.5in]{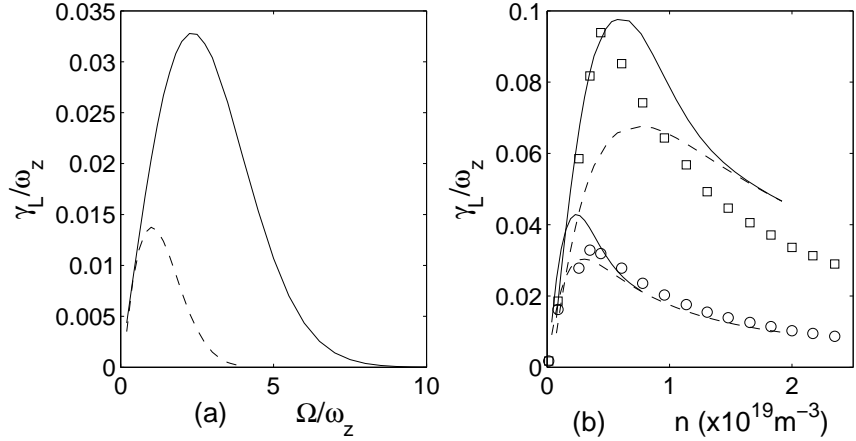} 
\caption{a) Landau damping rate of the dipole (dashed) and quadrupole (solid) modes as a function of the exchange frequency. b) Damping rates calculated by Nikuni {\it et al.}\cite{ref6} with a numerical solution of the kinetic equation for the dipole (circles) and quadrupole (squares) modes for $^{87}$Rb at $T=$800nK. Also shown are the damping rates calculated with the moments method\cite{ref6} (dashed), and with Landau damping included (solid).}
\label{fig:two}
\end{figure}

 There are two possible improvements to the model that immediately come to mind: 1) Better approximations of the functions $S^+_j(z)$ and the frequencies $\omega_j$ could be used. These could be obtained by including higher moments in the moments calculation and should give a more accurate value of the exchange field experienced  by spins, especially in orbits with $E>k_BT$.  2) The effects of collisions on Landau damping could be included. Although these might be rather difficult to incorporate within the framework of the current model, they might be easy to analyze for the homogeneous system, in Eqs.(1-3), for example, and perhaps the predictions of the trapped model could then be modified in a simple way. We leave this for future work.

\section*{ACKNOWLEDGMENTS}
This research was supported by NSF Grant DMR-0209606.

\end{document}